\documentclass[doublecol]{epl2} 
\usepackage{nicefrac}

\usepackage{times}
\usepackage{xcolor}
\usepackage{amsmath}
\usepackage{epsfig}
\usepackage{dcolumn}
\usepackage{amssymb}

\title{Nonequilibrium-DMFT based RIXS investigation of the two-orbital \mbox{Hubbard} model}

\author{Philipp Werner,\inst{1} Steven Johnston,\inst{2} and Martin Eckstein\inst{3}}
\shortauthor{Philipp Werner, Steven Johnston and Martin Eckstein}

\institute{                    
  \inst{1} Department of Physics, University of Fribourg, 1700 Fribourg, Switzerland\\
  \inst{2} Department of Physics and Astronomy, University of Tennessee, Knoxville, Tennessee 37996-1200, USA\\
  \inst{3} Department of Physics, University of Erlangen-Nuremberg, 91058 Erlangen, Germany
}

\abstract{
Resonant inelastic X-ray scattering (RIXS) detects various types of high- and low-energy elementary excitations in correlated solids, and this tool will play an increasingly important role in investigations of time-dependent phenomena in photo-excited systems. While theoretical frameworks for the computation of equilibrium RIXS spectra are well established, the development of appropriate methods for nonequilibrium simulations are an active research field. Here, we apply a recently developed nonequilibrium dynamical mean field theory (DMFT) based approach to compute the RIXS response of photo-excited two-orbital Mott insulators. The results demonstrate the feasibility of multi-orbital nonequilibrium RIXS calculations and the sensitivity of the quasi-elastic fluorescence-like features and $d$-$d$ excitation peaks on the nonequilibrium population of the Hubbard bands.    
}

\begin{document}

\maketitle

\section{Introduction}

Resonant inelastic X-ray scattering (RIXS) \cite{Luuk2011} is a powerful experimental tool for probing the charge \cite{Hill1998}, spin \cite{Braicovich2009,Schlappa2018}, orbital \cite{Chen2010,Schlappa2012} and lattice \cite{Ament2011,Johnston2016,Chaix2017} degrees of freedom in correlated solids. 
It is a photon-in, photon-out technique, where the incoming photon with energy $\omega_\text{in}$ and momentum $\vec{q}_\text{in}$ excites an electron from a core level into the conduction band. What is measured is the photon emitted with energy $\omega_\text{out}$ and momentum $\vec{q}_\text{out}$, when the excited electron or another one fills the core hole.  A recent development are time-resolved RIXS experiments \cite{Dean2016,Mitrano2019,Mitrano2020}, which track the RIXS signal in a material driven out of its equilibrium state by a laser pulse. Thanks to a high time resolution, these measurements can reveal the interplay between spin, orbital, and lattice degrees of freedom during the laser driving and relaxation dynamics.

The equilibrium RIXS amplitude $I$ can be obtained from cluster diagonalization and the Kramers-Heisenberg formula \cite{Kramers1925,Chen2019}
$I_{\vec{q}_\text{in}, \vec{q}_\text{out}}(\omega_\text{in}, \omega_\text{out})=\frac{1}{\pi}\text{Im}\langle\psi| (H_\text{cl} -E_0 -(\omega_\text{in}- \omega_\text{out})-i0_+)^{-1}|\psi\rangle$, 
with  
$|\psi\rangle = \sum_m e^{i(\vec{q}_\text{in}-\vec{q}_\text{out})\cdot \vec{r}_m} D^\dagger_{m}(H_\text{cl}-E_0-\omega_\text{in}-i\Gamma)^{-1} D_{m}|\phi_0\rangle.$
Here, $H_\text{cl}$ is the Hamiltonian of the cluster (including core levels), $E_0$ is the energy of the ground state $|\phi_0\rangle$, $D_{m}$ is the dipole transition operator for lattice site $m$, which can also depend on the photon polarization and electron spin, and $\Gamma$ is a parameter determining the lifetime of the core hole. These calculations capture the atomic multiplet structure very well, but provide only a crude description of the itinerant nature of the conduction electrons. 

The nonequilibrium generalization of this cluster based RIXS formalism has recently been formulated by Devereaux, Freericks and co-workers in Ref.~\cite{Chen2019}. It is based on the measurement of the four-point correlation function
\begin{align}
& S_{mn}(t_1^{\phantom\prime},t_2^{\phantom\prime},t_1',t_2')= \langle U(-\infty,t_1')D_{n}(t_1')U(t_1',t_2')D^\dagger_{n}(t_2')\nonumber\\
&\hspace{5mm} \times U(t_2',t_2^{\phantom\prime})D_{m}(t_2)U(t_2,t_1)D^\dagger_{m}(t_1)U(t_1,-\infty)\rangle, 
\label{4point}
\end{align}
with $U$ the time evolution operator, and has initially been demonstrated for a noninteracting problem, where this complicated correlation function can be explicitly calculated. The application of the method to interacting systems is numerically challenging, because of the difficulties of measuring and storing $S_{mn}(t_1^{\phantom\prime},t_2^{\phantom\prime},t_1',t_2')$. It has been successfully used to study a single-band Hubbard model with additional core-hole potential on a small cluster \cite{Wang2019}, but extensions of this approach to multi-orbital problems will be extremely costly. Moreover, while a small cluster may be sufficient for capturing the dynamics during the RIXS process, finding a finite cluster that can accurately represent the non-equilibrium dynamics  over a longer period of time after a pump laser excitation is challenging.

In a separate effort, a dynamical mean field theory (DMFT) \cite{Georges1996} based equilibrium RIXS formalism has been developed by Hariki, Kunes, and co-workers \cite{Hariki2018,Hariki2019}. In this scheme, the orbitals near the Fermi level are replaced by an impurity problem with a hybridization function derived from a DMFT calculation. 
The advantage of this approach is that it can treat the itinerant nature of the conduction electrons  
via the DMFT construction. It thus allows one to compute a fluorescence-like RIXS signal with a linear dependence of $\omega_\text{loss}=\omega_\text{in}-\omega_\text{out}$ on $\omega_\text{in}$. 
A current limitation of this method is that it does not provide information on the $\vec{q}$-dependence of the RIXS spectra. 

In Ref.~\cite{Eckstein2020} we showed how the DMFT based RIXS approach can be implemented on the real-time axis, and thereby extended to nonequilibrium problems. The basic idea is to simulate the RIXS probe pulse explicitly using nonequilibrium DMFT \cite{Aoki2014}, and to measure a time-dependent correlation function from which the number of emitted photons can be obtained. We will use here an implementation without rotating wave approximation, and with the noncrossing approximation (NCA) \cite{Keiter1971,Eckstein2010} as impurity solver (for details, see Ref.~\cite{Eckstein2020}), to investigate the equilibrium and nonequilibrium RIXS spectra of the half-filled an quarter-filled two-orbital Hubbard model in the Mott insulating regime. In particular, we will show how these spectra are modified by a short but strong photo-doping pulse.  

\section{Model and method}

We consider a two-orbital Hubbard model with a coupling to core levels. The local (atomic) Hamiltonian reads
\begin{align}
H_\text{loc}=&\sum_{\alpha} U n_{\alpha\uparrow}n_{\alpha\downarrow} \nonumber\\
&+ \sum_{\sigma}[(U-2J) n_{1\sigma}n_{2\bar\sigma}+(U-3J)n_{1\sigma}n_{2\sigma}]\nonumber\\
&+ U_{cd} n_cn_d+\tfrac\Delta2(n_d-2n_c)-\mu(n_d+n_c),
\label{ham}
\end{align}
where $\alpha=1,2$ denotes the orbitals, $\sigma=\uparrow,\downarrow$ denotes spin, $d_{\alpha\sigma}$ and $n_{\alpha\sigma}$ are the corresponding annihilation and density operators, $c_\sigma$ and $n_{c\sigma}$ are the annihilation and density operators for the core level, $U$ is the intra-orbital interaction, $J$ is the Hund coupling, $U_{cd}$ is the interaction with the core level, and $\mu$ is the chemical potential.  We use the notations $n_d=\sum_{\alpha,\sigma}n_{\alpha\sigma}$,  $n_c=\sum_\sigma n_{c\sigma}$ and introduce a parameter $\Delta$ to shift the core energy relative to the $d$ levels. To mimic the probe pulse, we add a dipolar coupling between the core level and the first $d$-orbital: $H_\text{probe}=E_\text{probe}(t)\sum_\sigma (d^\dagger_{1\sigma}c^{\phantom\dagger}_\sigma + h. c.)$, with a field of the form $E_\text{probe}(t)=E_\text{probe}f_\text{probe}(t-t_\text{probe})\sin(\omega_\text{in}(t-t_\text{probe}))$ whose amplitude $E_\text{probe}$ is weak enough that the RIXS signal is quadratic in $E_\text{probe}$, $\omega_\text{in}$ the incoming frequency of the order of the $c$--$d$ energy splitting, and $f(t-t_\text{probe})$ an envelope function centered at time $t_\text{probe}$. The $d$-electrons hop between lattice sites with the kinetic term $H_\text{kin}=\sum_{\langle i,j\rangle,\alpha,\sigma} v(t)(d^\dagger_{i\alpha\sigma}d^{\phantom\dagger}_{j\alpha\sigma}+ h.c.)$. A time-dependent modulation of the hopping parameter $v$ with frequency $\omega_\text{pump}$ of the order of $U$, amplitude $v_\text{pump}$ and envelope $f_\text{pump}$ will be used to ``photo-excite" the system: $v(t)=v_0+v_\text{pump}f_\text{pump}(t-t_\text{pump})\sin(\omega_\text{pump}(t-t_\text{pump}))$. Furthermore, to mimic the finite core-hole lifetime (parameter $\Gamma$ in the Kramers-Heisenberg approach), we couple a fermionic bath with a box-shaped density of states (DOS) to the core level. 

We consider an infinite-dimensional Bethe lattice with appropriately renormalized hopping amplitude, so that the DMFT self-consistency gives a direct relation between the impurity hybridization function $\Lambda_{\nu,\mu,\sigma}$ and the local Green's functions $G_{\mu,\nu,\sigma}$ (with $\nu,\mu=(1,2,c)$) \cite{Georges1996}:
${\bf \Lambda_\sigma}(t,t') = {\bf V}(t) {\bf G_\sigma}(t,t') {\bf V}(t')$.
Here, the bold symbols represent $3\times 3$ matrices. We choose $V_{11}(t)=V_{22}(t)=v(t)$ and $V_{cc}(t)=0$ (localized core level). The hybridization function ${\bf \Lambda_\sigma}$ of the (equilibrium or photo-excited) lattice system represents the environment of the probed site and is not modified in the simulation of the RIXS process. The RIXS probe pulse, however, introduces coherence between the $c$ and $d_1$ orbitals, and produces nonzero off-diagonal elements in ${\bf G_\sigma}$.  

The RIXS signal can be computed from the correlation functions $D_{\alpha,\sigma,\sigma'}(t,t')=-i\langle T_\mathcal{C} P^{\phantom\dagger}_{\alpha\sigma}(t)P_{\alpha\sigma'}^\dagger(t')\rangle$ with $P_{\alpha\sigma}=c^\dagger_\sigma d^{\phantom\dagger}_{\alpha\sigma}$, see Ref.~\cite{Eckstein2020} for details. In the following, we will focus on the signal emitted by orbital $\alpha=1$, and use $v_0=1$ as the unit of energy ($\hbar/v_0$ as the unit of time), which corresponds to a noninteracting lattice DOS of bandwidth $4$.

\section{Results} 

\begin{figure*}[t]
\begin{center}
\includegraphics[angle=-90, width=0.33\textwidth]{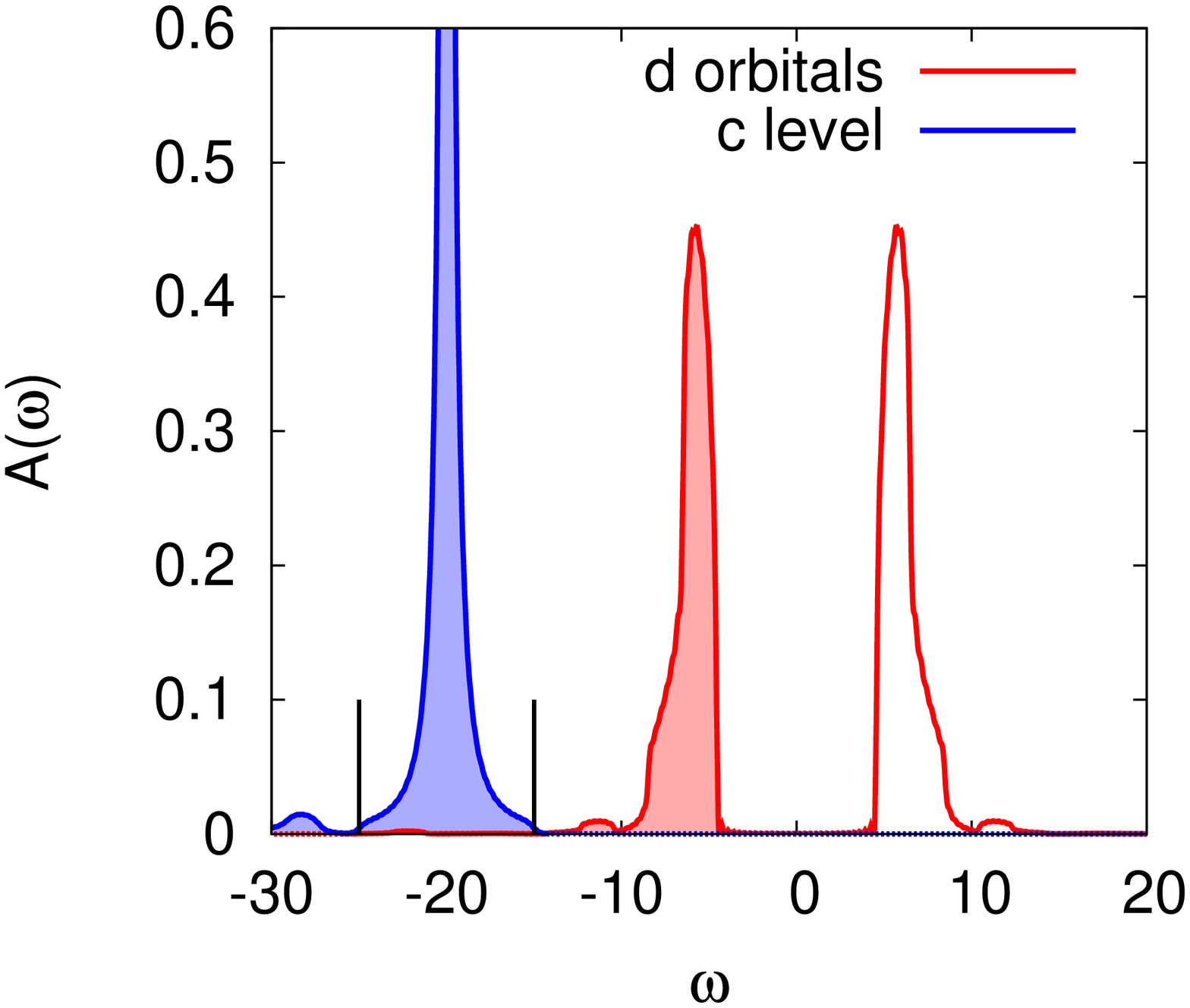} 
\includegraphics[angle=-90, width=0.33\textwidth]{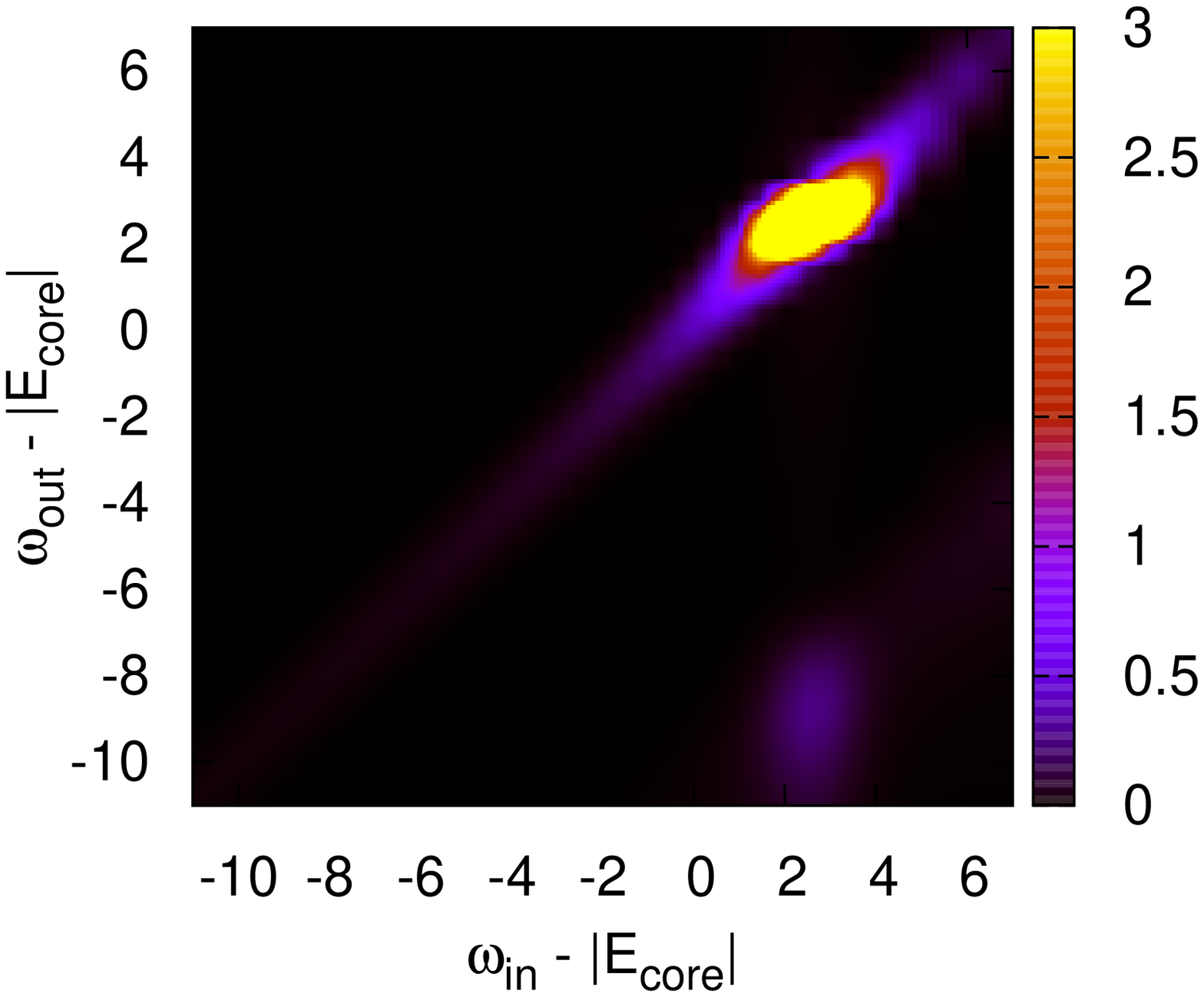} 
\includegraphics[angle=-90, width=0.33\textwidth]{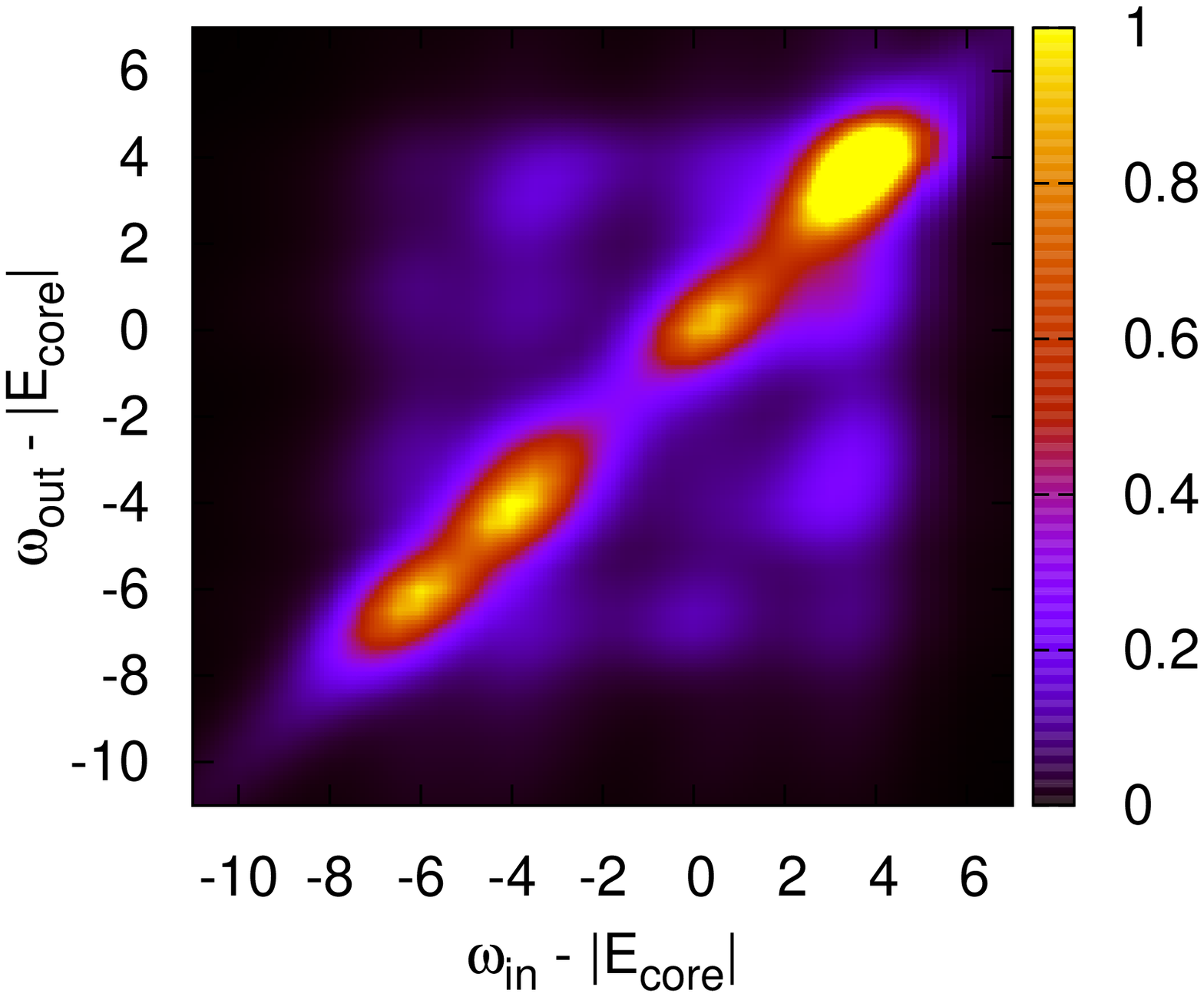} 
\caption{
Left panel: equilibrium spectral functions of the half-filled model with $U=10$, $J=2$, $U_{cd}=3$, $E_\text{core}=-20$. The coupling to a Fermion bath with box-shaped DOS from $-25$ to $-15$ (black lines) and coupling strength $v_\text{bath}=1$ \cite{Eckstein2020} leads to a broadening of the $c$-DOS. The main Hubbard bands in the $d$-DOS are at $\pm 6$. The middle panel shows the local equilibrium RIXS response, normalized by $E_\text{probe}^2$, and the right panel the normalized local RIXS signal of the photo-doped system with $\omega_\text{pump}=12$ and $t_\text{probe}-t_\text{pump}\approx 2$.  
}
\label{fig_half}
\end{center}
\end{figure*}

{\it Half-filled system.} We first consider a half-filled two-orbital Hubbard model with $U=10$, $J=2$, and $U_{cd}=3$. Because of the Hund coupling, the dominant atomic configurations in equilibrium are the half-filled high-spin doublon states with one electron in each orbital and parallel spins.  The splitting between the Hubbard bands is $U+J$. We choose the parameters $\mu$ and $\Delta$ such that (i) the energies  for $d$-electron addition and removal (upper and lower Hubbard bands) are symmetric at $\pm 6$, and (ii) the core removal energy is $E_\text{core}=-20$ \cite{footnote}.  For model (\ref{ham}), condition (i) implies that the electron addition energy is $E(3,1)-E(2,2)=2U-2J+2U_{cd}+\tfrac{\Delta}{2}-\mu\equiv 6$, while (ii) implies $E(2,1)-E(2,2)=2U_{cd}-\Delta-\mu\equiv -20$. Here, $E(n_d,n_c)$ is the energy of the corresponding state with filling $n_d$ and $n_c$ for the $d$ and $c$ orbitals, respectively. These two conditions fix $\Delta=6.667$ and $\mu=19.33$. The corresponding equilibrium spectral functions are shown in the left panel of Fig.~\ref{fig_half}. Apart from the main Hubbard bands, we recognize shoulder and satellite structures at higher energies, which are associated with local spin excitations (triplon insertion plus hopping to a neighboring site, which leaves behind a low-spin doublon \cite{Lysne_2020}).

The middle panel shows the equilibrium RIXS signal $I_\text{RIXS}$ extracted from the $D_{1,\sigma,\sigma'}$ correlation functions, summed over the spin channels, and normalized by $E_\text{probe}^2$. The probe pulse has an envelope $f_\text{probe}(t)=\exp(-t^2/2)$, corresponding to a width of $\Delta t\approx 2$ (and hence to an energy resolution of about $\Delta E \approx \tfrac{1}{4}$), and an amplitude $E_\text{probe}=0.05$, which is well within the perturbative regime where $I_\text{RIXS}\propto E_\text{probe}^2$. The signal exhibits a single strong peak at $\omega_\text{in} -|E_\text{core}|\approx 3$, which is associated with processes where a core electron is excited into the upper Hubbard band and decays back. Without the core-valence interaction $U_{cd}$, the corresponding energy  would be given by the electron insertion into the dominant subband of the upper Hubbard band at  $\omega_\text{in}-|E_\text{core}|\approx \tfrac{U+J}{2}=6$, but because of the core hole interaction, it is shifted to $\approx \tfrac{U+J}{2}-U_{cd} = 3$. 
Although it is not very clear on the scale of the plot, the signal peaks at  a value of $\omega_\text{out}$ which is almost independent of $\omega_\text{in}$ within an energy range approximately equal to the width of the Hubbard band. This implies that $\omega_\text{loss}=\omega_\text{in}-\omega_\text{out}$ is approximately proportional to $\omega_\text{in}$ in the corresponding energy range, i.e. kinetic energy is transmitted to other electrons during the RIXS process (fluorescent line). In addition, we see a very weak signal at $\omega_\text{out}-|E_\text{core}|\approx -9$, corresponding to $\omega_\text{loss}\approx U+J$. This signal originates from a RIXS process that leaves a singlon-triplon pair in the system after the core hole is filled ($d$-$d$ excitation). The energy of $U+J$ corresponds to the energy difference $E(3,2)+E(1,2)-2E(2,2)$ between two sites with a singlon-triplon pair and two sites in their ground state.

\begin{figure}[t]
\begin{center}
\includegraphics[angle=-90, width=0.8\columnwidth]{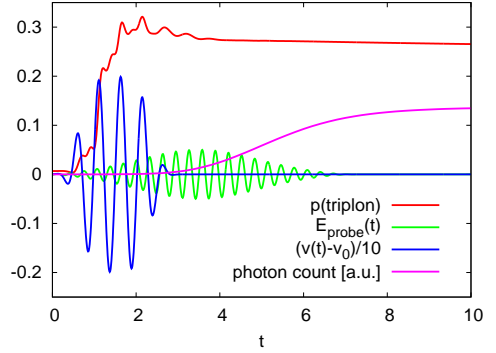}
\caption{
Evolution of the density of triplons (red line) induced by a short hopping modulation pulse (blue line). The state of the system immediately after the pulse is probed by a RIXS pulse (green line, here for $\omega_\text{in}=|E_\text{core}|$). The violet curve is proportional to the photon count (for $\omega_\text{out}=|E_\text{core}|$).
}
\label{fig_doublon_half}
\end{center}
\end{figure}

The right panel of Fig.~\ref{fig_half} shows  $I_\text{RIXS}/E_\text{probe}^2$ of the photo-doped system. Here, a few-cycle hopping modulation is applied to generate a substantial density of singlons and triplons, before the RIXS spectrum is probed.  Figure~\ref{fig_doublon_half} shows the evolution of the triplon probability, 
$p(\text{triplon})=\langle n_{1\uparrow}n_{1\downarrow} n_2 + n_1 n_{2\uparrow}n_{2\downarrow}-4n_{1\uparrow}n_{1\downarrow}n_{2\uparrow}n_{2\downarrow}\rangle$, with  $n_\alpha=n_{\alpha\uparrow}+n_{\alpha\downarrow}$, 
which is initially very low, because the system is predominantly in half-filled high-spin states with one electron per orbital. During the hopping modulation pulse with frequency $\omega_\text{pulse}=U+J$ (blue line), the triplon population increases even beyond the infinite temperature value of $\tfrac14$, while the probability of high-spin doublons decreases from $0.99$ to $0.17$. Due to particle-hole symmetry, there is a corresponding increase in the density of singlons. The RIXS probing pulse, which partially overlaps with the pump pulse, measures the system immediately after the pump excitation (time delay $t_\text{probe}-t_\text{pump}\approx 2$). We have also measured the RIXS spectrum after a longer time delay $t_\text{probe}-t_\text{pump}\approx 4$, and obtained an almost identical result. This can be explained by the fact that the transfer of kinetic energy to local spin excitations occurs quickly (during the pump pulse) \cite{Strand_2017}, while the life-time of the singlons and triplons in this large-gap insulator is orders of magnitude longer than our simulation times \cite{Eckstein_2011}.     

As we can see in the right panel of Fig.~\ref{fig_half}, the photodoping generates several prominent new peaks in $I_\text{RIXS}$ with $\omega_\text{loss}\approx 0$ (diagonal), and several weaker peaks corresponding to $d$-$d$ excitations with $\omega_\text{loss}>0$ (lower-right triangle) and to $d$-$d$ de-excitations with $\omega_\text{loss}<0$ (upper-left triangle). The presence of high- and low-spin doublons in the photo-doped system now leads to three quasi-elastic features associated with doublon-to-triplon excitations, at energies $E(3,1)-E(2,2)=3$, $3-J=1$, $3-3J=-3$. 
The new quasi-elastic peaks at $\omega_\text{in,out}-|E_\text{core}|\approx -6$, $-4$ and $0$ correspond to RIXS processes where an electron is excited from the core level to a singlon, and eventually decays back. The peak near $-6$ can be associated with an intermediate high-spin doublon state (e.~g. \mbox{$|\!\uparrow,\uparrow\rangle$}), the peak near $-4$ with an intermediate low-spin doublon state with one electron in each orbital (e.~g. \mbox{$|\!\uparrow,\downarrow\rangle$}), and the peak near $0$ with an intermediate doublon state with a doubly occupied and an empty orbital (e.~g. \mbox{$|0,\uparrow\downarrow\rangle$}). Indeed, since the $U_{cd}$- and $\mu$-contributions to the local energy do not change in these processes, the corresponding excitation energies $E(2,1)-E(1,2)$ are $U-3J+\tfrac{3}{2}\Delta=14$,  $U-2J+\tfrac{3}{2}\Delta=16$ and $U+\tfrac{3}{2}\Delta=20$, which upon subtraction of $|E_\text{core}|$ yields the observed peak positions. 
The excitation of a core electron to a triplon state produces a change in the $U_{cd}$ contribution, and the corresponding excitation energy $E(4,1)-E(3,2)$ becomes $3U-5J-2U_{cd}+\tfrac32\Delta=24$. Hence, these excitations explain the prominent peak at $\omega_\text{in,out}-|E_\text{core}|\approx 4$, which partially overlaps with the peak at $\approx 3$ corresponding to excitations to doublon states. The latter is substantially weaker in the photo-doped system than in equilibrium, because the high-spin doublon states no longer dominate. 
There is also a small density of empty sites in the photo-doped state, which results in a weak quasi-elastic feature near $E(1,1)-E(0,2)=13$, i.e. $\omega_\text{in,out}-|E_\text{core}|\approx -7$. We summarize the different quasi-elastic features in the half-filled equilibrium and photo-doped systems in Tab.~\ref{tab}.

\begin{table}
\begin{center}
\begin{tabular}{l|c|c}
relevant transitions & half-filled & quarter-filled  \\
\hline
equilibrium:& &  \\
$(2^h,2)\leftrightarrow (3,1)$ & 3 & \\
$(1,2)\leftrightarrow ({\bf 2},1)$ & & $-1,1,5$ \\
\hline
photo-doped:&  &  \\
$(0,2)\leftrightarrow (1,1)$ & $-7$ & $-2$ \\
$(1,2)\leftrightarrow ({\bf 2},1)$ & $-6,-4,0$ & $-1,1,5$ \\
$({\bf 2},2)\leftrightarrow (3,1)$ & $-3,1,3$ & $2,6,8$ \\
$(3,2)\leftrightarrow (4,1)$ & $4$ & $9$ \\ 
\hline
\end{tabular}
\end{center}
\caption{Energies $\omega_\text{in,out}-|E_\text{core}|$ of the quasi-elastic features in the half-filled and quarter-filled system. The left (right) number in the brackets indicates the filling of the $d$ ($c$) orbital. $2^h$ denotes the high-spin doublon state, and ${\bf 2}$ the three different types of doublons.}
\label{tab}
\end{table}

\begin{figure*}[ht]
\begin{center}
\includegraphics[angle=-90, width=0.33\textwidth]{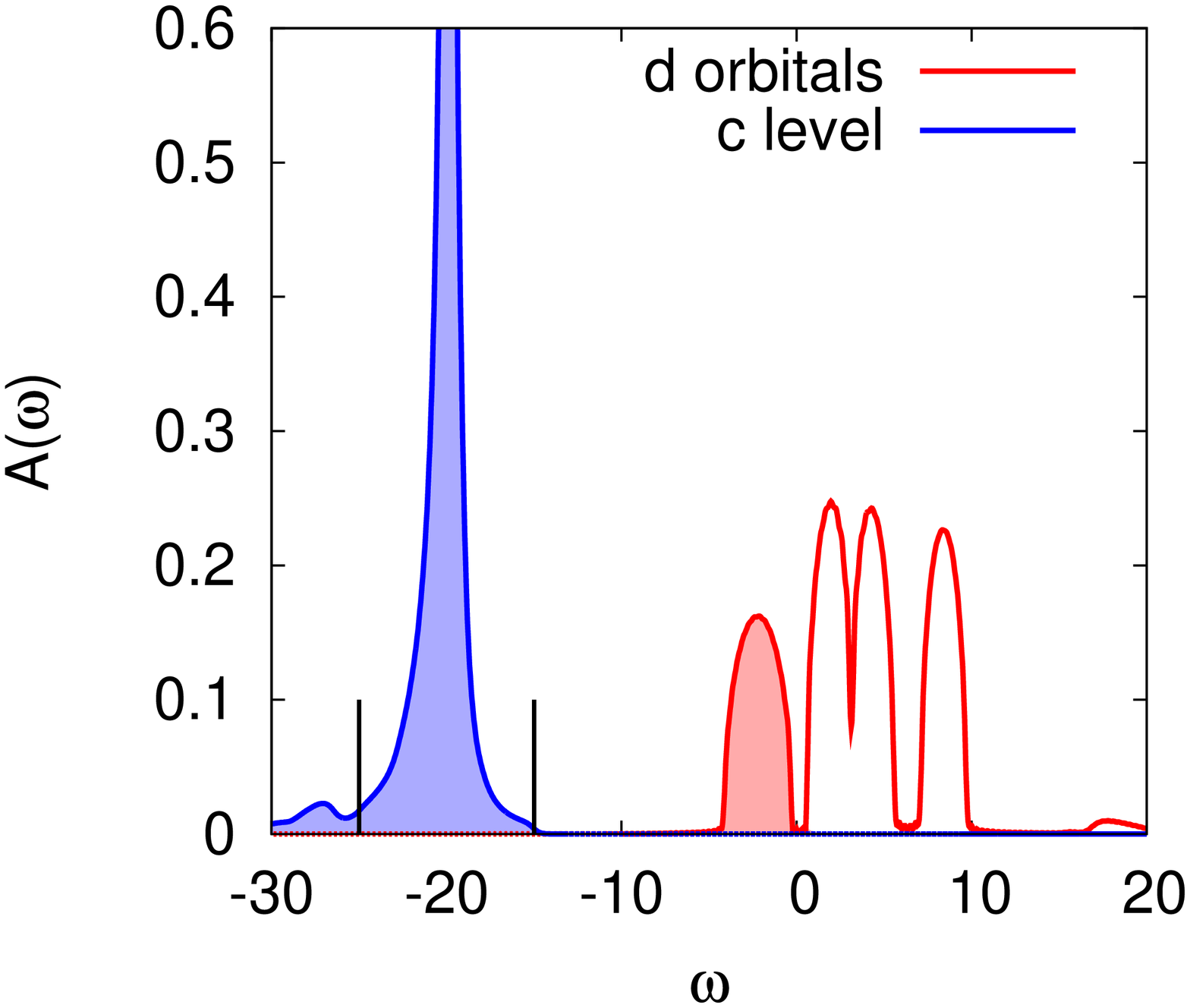} 
\includegraphics[angle=-90, width=0.33\textwidth]{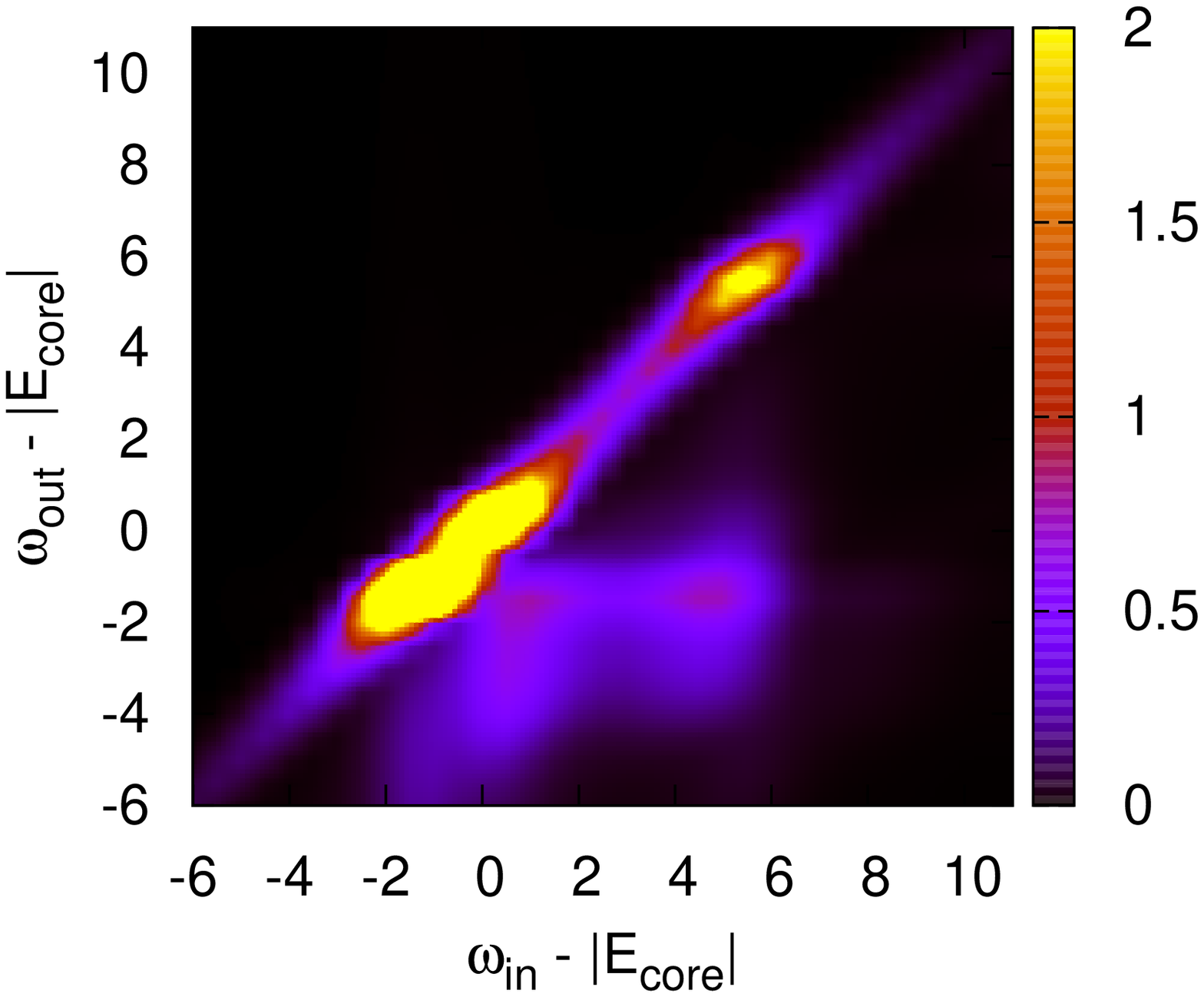} 
\includegraphics[angle=-90, width=0.33\textwidth]{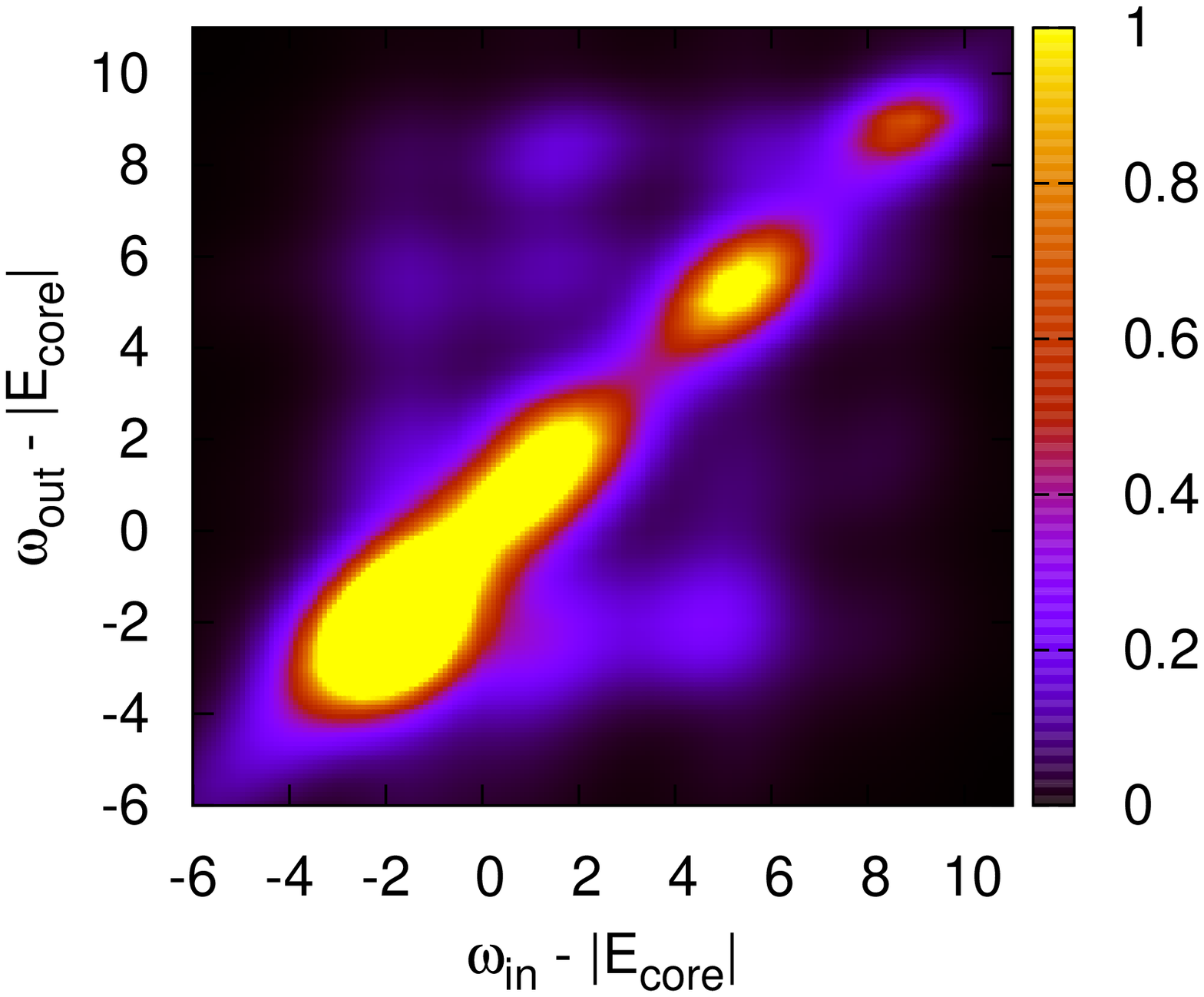} 
\caption{
Left panel: equilibrium spectral functions of the quarter-filled model with $U=10$, $J=2$, $U_{cd}=3$, $E_\text{core}=-20$. The coupling to a Fermion bath with box-shaped DOS from $-25$ to $-15$ (black lines) and coupling strength $v_\text{bath}=1$ \cite{Eckstein2020} leads to a broadening of the $c$-DOS. The lower Hubbard band (singly occupied sites) in the $d$-DOS is at $-2$, while the upper Hubbard band (doubly occupied sites) has three subbands at energies $2$, $4$ and $6$.  The middle panel shows the local equilibrium RIXS response, normalized by $E_\text{probe}^2$, and the right panel the normalized local RIXS signal of the photo-doped system with $\omega_\text{pump}=10$ and $t_\text{probe}-t_\text{pump}\approx 2$.  
}
\label{fig_quarter}
\end{center}
\end{figure*}   

The weaker RIXS features on the $\omega_\text{loss}>0$ side (lower-right triangle) can be associated with processes involving the neighboring sites. For example the small peak at $\omega_\text{in}-|E_\text{core}|\approx -4$, $\omega_\text{out}-|E_\text{core}|\approx -6$ can be explained by a process where a singlon is converted into a low-spin doublon with interaction energy $U-2J$, which is then (via a second-order hopping process) flipped into a high-spin doublon with energy $U-3J$, before an electron decays to fill the core hole:
\mbox{$(|0,\downarrow\rangle|\!\uparrow\downarrow\rangle)$}$\xrightarrow{\omega_\text{in}}
(|\!\!\uparrow,\downarrow\rangle|\!\!\downarrow\rangle)\xrightarrow{\text{ex}}
(|\!\!\uparrow,\uparrow\rangle|\!\!\downarrow\rangle)\xrightarrow{\omega_\text{out}}
(|0,\uparrow\rangle|\!\!\uparrow\downarrow\rangle)$,
where the configurations $(|d\rangle|c\rangle)$ indicate the valence $(|d\rangle)$ and core $(|c\rangle)$ configurations on the probe site. The energy loss $\omega_\text{loss}\approx J$ is either converted into a local spin excitation on the neighboring site, or converted into kinetic energy. Similarly, the feature at $\omega_\text{in}-|E_\text{core}|\approx 0$, $\omega_\text{out}-|E_\text{core}|\approx -6$ corresponds to an intermediate state with energy $U$ (doublon with both electrons in the same orbital). The much weaker feature at $\omega_\text{in}-|E_\text{core}|\approx 0$, $\omega_\text{out}-|E_\text{core}|\approx -4$ indicates that the states involved in the corresponding second-order processes are less populated or characterized by weaker exchange couplings. 

$d$-$d$ excitation peaks at $\omega_\text{in}-|E_\text{core}|\approx 3$ are related to excitations of a core electron to a high-spin doublon state (see Tab.~\ref{tab}). The resulting triplon can hop to a neighboring site, leaving behind doublons in various configurations. The decay to a singlon plus filled core hole then produces photons at $\omega_\text{out}-|E_\text{core}|\approx 0$, $-4$, $-6$. 
In the photo-doped state, the intermediate step (hopping of triplon) is aided by the presence of low-spin doublons on neighboring sites, which reduces the amount of kinetic energy that needs to be absorbed or emitted. 
Alternatively, the triplon can change its configuration in a second-order process and then decay to a low-spin doublon plus a filled core hole. In this case, the emitted energy peaks around $\omega_\text{out}-|E_\text{core}|\approx 3-J=1$ and $\approx 3-3J=-3$, while the intermediate step is again aided by the presence of neighboring low-spin doublons. The superposition of these two series of peaks explains the observed loss features. Note that the peak at $\omega_\text{out}-|E_\text{core}|=-9$, which appeared in the equilibrium spectrum, becomes very weak in the photo-doped state. This is because pairs of neighboring high-spin doublons are unlikely in the photo-doped system. 

We next turn our attention to the features in the upper left triangle, corresponding to $\omega_\text{loss}<0$. In a photo-doped system, energy can be gained in processes involving singlon-triplon recombination or local state transitions. The gain features at $\omega_\text{in}-|E_\text{core}|\approx -6$ and $\omega_\text{out}-|E_\text{core}|\approx-4$, $0$, as well as  $\omega_\text{in}-|E_\text{core}|\approx -4$ and  $\omega_\text{out}-|E_\text{core}|\approx 0$ can be explained as energy absorptions of $J$, $2J$ and $3J$ from local state transitions in second order hoppings between neighboring doublons. These become possible because low-spin doublons are already present before the RIXS pulse. The strong feature near $\omega_\text{out}-|E_\text{core}|\approx 3$ and  $\omega_\text{in}-|E_\text{core}|$ between $-4$ and $-3$ originates from emission processes that leave behind a high-spin doublon state. The initial state is either a doublon with local energy $U$ ($\omega_\text{in}-|E_\text{core}|\approx -3$) or a singlon. In the latter case, the doublon state after the RIXS excitation has local energy $U-2J$ (hence $\omega_\text{in}-|E_\text{core}|\approx -4$), and before the de-excitation, a triplon hops in from a neighboring site:
$(|0,\downarrow\rangle|\!\!\uparrow\downarrow\rangle)\xrightarrow{\omega_\text{in}}
(|\!\!\uparrow,\downarrow\rangle|\!\!\downarrow\rangle)\xrightarrow{\text{hop}}
(|\!\!\uparrow\downarrow,\uparrow\rangle|\!\!\downarrow\rangle)\xrightarrow{\omega_\text{out}}
(|\!\!\downarrow,\downarrow\rangle|\!\!\uparrow\downarrow\rangle)$.
Again, such processes become possible because singlons and triplons are present with high probability in the photo-doped system.  

{\it Quarter-filled system.} As a second example, we consider the quarter-filled two-band model with $U=10$, $J=2$, $U_{cd}=3$, $E_\text{core}=-20$, which has a very different equilibrium DOS, as illustrated in the left panel of Fig.~\ref{fig_quarter}. Here, in order to place the core level at $-20$, and the highest energy subband of the upper Hubbard band at $8$, we choose the electron removal and addition energies as $-\mu+U_{cd}-\Delta\equiv -20$ and $-\mu+2U_{cd}+U+\tfrac{\Delta}{2}\equiv 8$, which gives $\Delta=10$ and $\mu=13$.  With this choice, the lower Hubbard band (representing singly occupied states) is centered at $\approx -2$, and the subbands of the upper Hubbard band (corresponding to high-spin and low-spin doublon states) at $\approx 2$, $4$, and $8$, see left panel of Fig.~\ref{fig_quarter}. In the atomic limit, the energy of creating a high-spin doublon becomes $E(2,1)-E(1,2)=U-3J+\tfrac32\Delta=19$, and that of creating a low-spin doublon with electrons in different orbitals (the same orbital) $U-2J+\tfrac32\Delta=21$ ($U+\tfrac32\Delta=25$). Hence, in the equilibrium RIXS spectrum we expect dominant features around $\omega_\text{in/out}-|E_\text{core}| \approx -1$, $1$, $5$, which are associated with the creation and decay of these different types of doublons. This is indeed approximately the case as shown in the middle panel of Fig.~\ref{fig_quarter}. Again, the horizontal elongation of these dominant features indicates energy loss through scattering within a given Hubbard-subband, and hence a fluorescence-like behavior. 

Because the energy gaps are smaller than in the half-filled case, prominent $d$-$d$ excitation peaks are visible already in the equilibrium RIXS spectrum. These energy loss features appear near $\omega_\text{out}-|E_\text{core}|\approx -2$ and $-1$. The signal at $-2$ can be associated with the transition from a singly occupied $d$ site with core hole to an empty $d$ site with filled core hole. This type of emission occurs in processes that leave behind a doublon-hole pair in the singly occupied background: $(1,2)(1,2) \xrightarrow{\omega_\text{in}} (2,1)(1,2) \xrightarrow{\text{hop}} (1,1)(2,2) \xrightarrow{\omega_\text{out}} (0,2)(2,2)$. Here, the two brackets correspond to neighboring sites, the first entry to the $d$ occupation and the second entry to the $c$ occupation. The first arrow represents the excitation by the probe pulse, the last arrow the de-excitation, and the middle arrow a hopping process, which costs an energy $\Delta E=U_{cd}-nJ$,  where $n$ depends on the specific doublon configuration. In the case of the signal at $-2$, this energy is absorbed from the system (e.~g. from the kinetic energy of the carriers) and does not show up in $\omega_\text{out}$. If on the other hand, we combine the last two processes above into a single ``hopping plus emission" process, the energy $\Delta E$ must be subtracted from $\omega_\text{out}$, which explains the weaker signals at $\approx -2-U_{cd}=-5$ and $-2-(U_{cd}-J)=-3$. 
Photons with $\omega_\text{out}-|E_\text{core}|\approx -1$ are emitted by the de-excitation of a high-spin doublon. For example, the peak around $\omega_\text{in}-|E_\text{core}|\approx 5$ and $\omega_\text{out}-|E_\text{core}|\approx -1$ originates from a RIXS process where the incoming pulse creates a low-spin doublon with interaction energy $U$, which in an intermediate step is converted into a high-spin doublon. This intermediate step releases an energy $\Delta E = 3J = 6$, which is converted into kinetic energy or additional doublon-holon excitations via the Hund-excitation analogue of impact ionization \cite{Petocchi_2019}. 

\begin{figure}[t]
\begin{center}
\includegraphics[angle=-90, width=1\columnwidth]{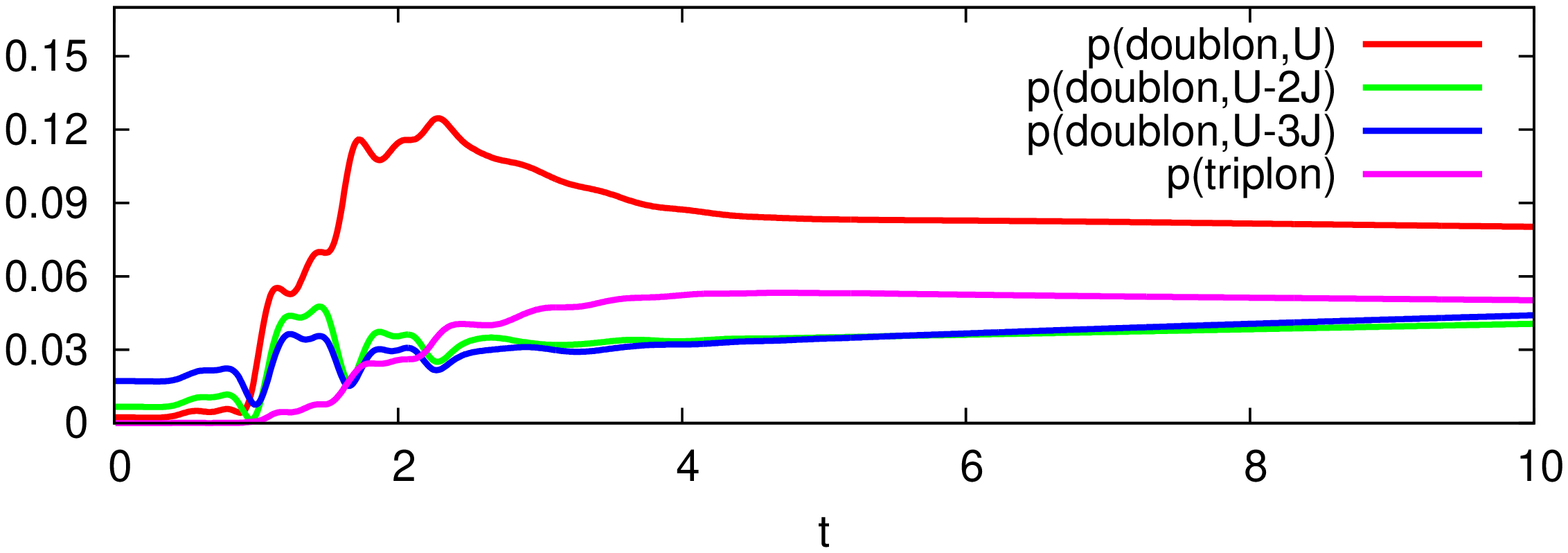}\\ 
\includegraphics[angle=-90, width=0.49\columnwidth]{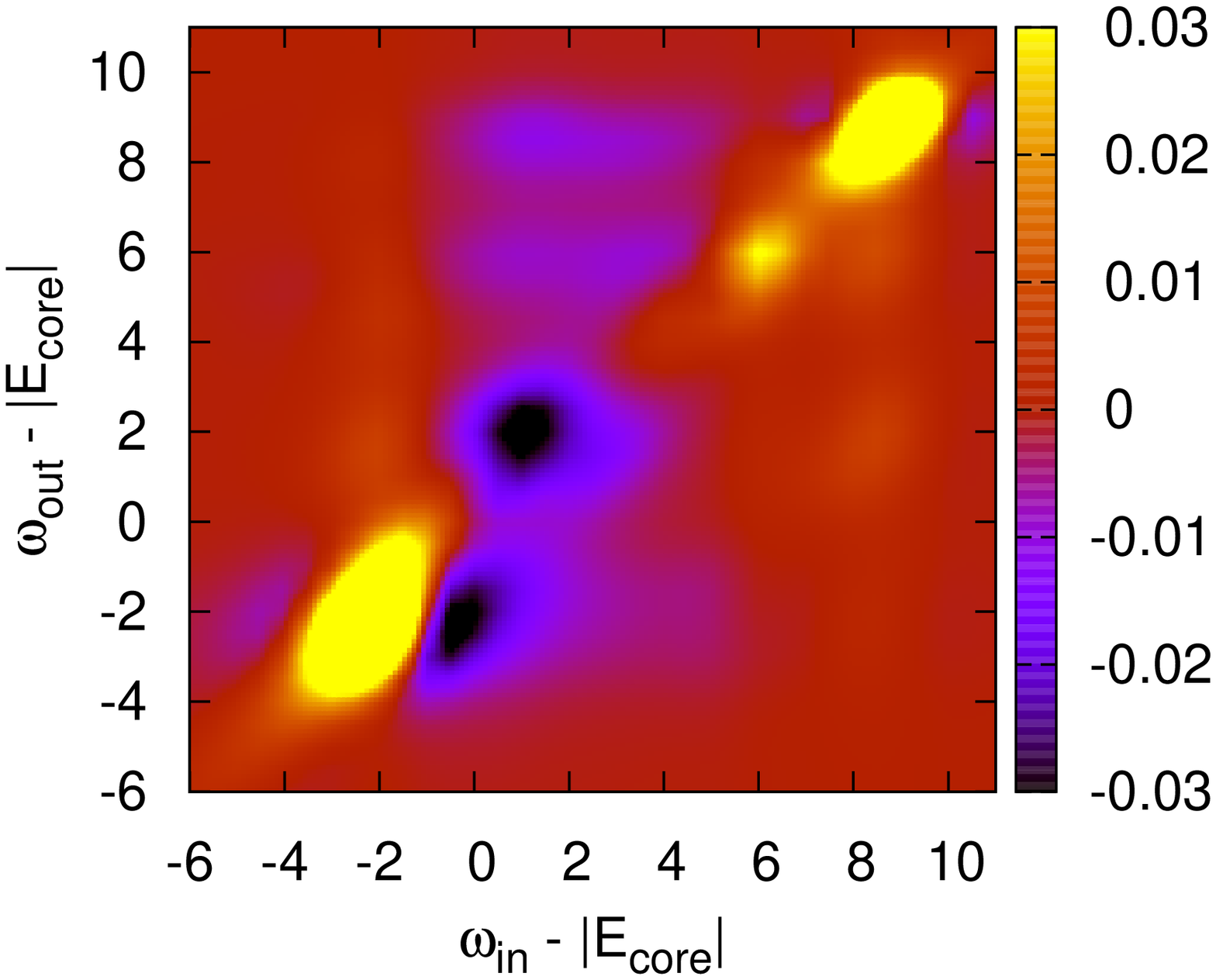} 
\hfill
\includegraphics[angle=-90, width=0.49\columnwidth]{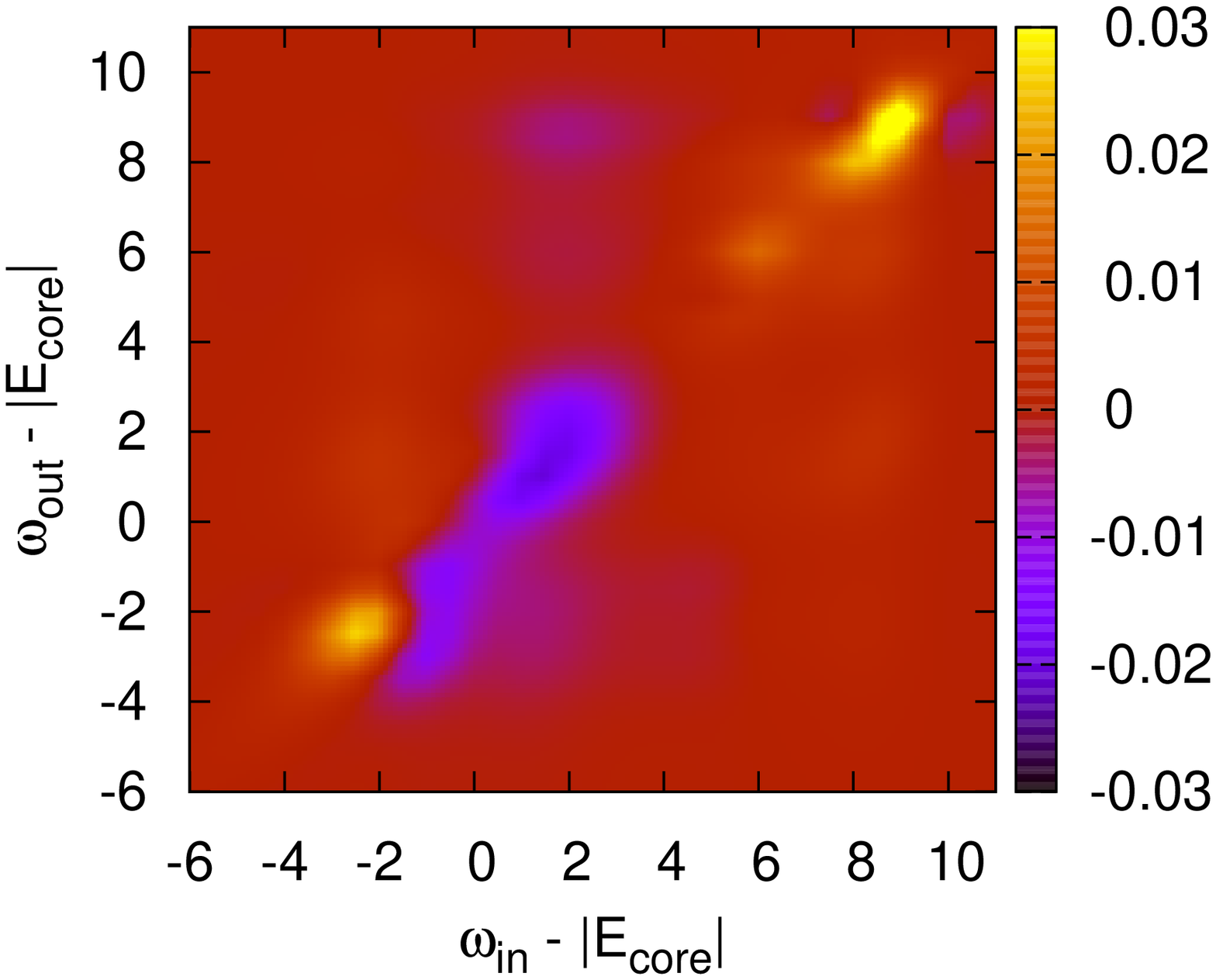} 
\caption{
Top panel: Time evolution of the doublon and triplon probabilities in the photo-doped quarter-filled system. Here, we label the type of doublon by the corresponding local energy. Bottom panels: 
Time dependent change in $I_\text{RIXS}/E_\text{probe}^2$. The left panel shows the difference for probe pulses centered at $t_\text{probe}=4.5$ and $3.5$ and the right panel the difference for probe pulses centered at $t_\text{probe}=5.5$ and $4.5$.
}
\label{fig_quarter_diff}
\end{center}
\end{figure}

The right hand panel of Fig.~\ref{fig_quarter} shows the RIXS spectrum after a photo-doping pulse. Here, we use the same set-up as illustrated in Fig.~\ref{fig_doublon_half}, but with a pump pulse frequency $\omega_\text{pump}=U$. As in the case of the half-filled system, the photo-doping results in additional peaks on the diagonal (quasi-elastic processes), as well as new gain and loss features in the upper left and lower-right triangle. The most prominent new peak appears at $\omega_\text{in,out}\approx -2$. This feature corresponds to the excitation of a core electron to an initially empty site, and the decay of this electron back to the core. Indeed, the corresponding excitation energy is $E(1,1)-E(0,2)=U_{cd}+\tfrac{3}{2}\Delta=18$, so that $\omega_\text{in,out}-|E_\text{core}|=-2$. There are also doublons in the initial photo-doped state. Electron addition to a high-spin doublon yields $E(3,1)-E(2,2)=2U-2J-U_{cd}+\tfrac{3}{2}\Delta=28$. Hence, three features on the diagonal at energies $\omega_\text{in,out}-|E_\text{core}|\approx 8$, $8-2J=6$, $8-3J=2$ can be associated with the creation and decay of triplons (Tab.~\ref{tab}). The prominent peak near $\omega_\text{in,out}-|E_\text{core}|\approx 9$ has  a different origin: it can be associated with the creation and decay of quadruplons (addition of a core electron to a triplon), since $E(4,2)-E(3,2)=3U-5J-2U_{cd}+\tfrac32\Delta=29$. Indeed, our photo-doping pulse is strong enough that it not only creates doublons, but also a non-negligible density of triplons, see the top panel of Fig.~\ref{fig_quarter_diff}. 

The features associated with $d$-$d$ excitations ($\omega_\text{loss}>0$) in the equilibrium spectrum get weaker in the photo-doped state, because the corresponding initial states (singlons) are less probable, and the production of doublon-hole pairs is suppressed if there is already a high density of doublons and holons in the system. A closer inspection further reveals that the dominant loss peak near $\omega_\text{in}-|E_\text{core}|\approx 1$ seems to have shifted to a lower energy. We interpret this peak as a holon-assisted feature. The RIXS pulse creates a low-energy doublon, which is converted into a singlon with the help of a neighboring holon. In this case, the intermediate step releases $\Delta E = U_{cd}-(U-3J)=-3$, which can be converted into kinetic energy or Hund excitations. The filling of the core hole from a singlon then emits light with $\omega_\text{out}-|E_\text{core}|\approx -2$.

Gain features with $\omega_\text{loss}<0$ can be associated with doublon or triplon assisted intermediate processes. For example, the peak at $\omega_\text{in}-|E_\text{core}|\approx -2$ and $\omega_\text{out}-|E_\text{core}|\approx 5$ corresponds to the creation of a singlon on an empty site by the RIXS pulse. This singlon is then converted into a low-spin same-orbital doublon through electron exchange with a neighboring doublon. This doublon subsequently decays into a singlon by filling the core hole. Energy is gained in this process from the recombination of a pump-induced doublon-hole pair. There are additional signals at the same $\omega_\text{out}$ from processes with $\omega_\text{in}-|E_\text{core}|\approx -1$ and $1$ (one of them overlapping with the former peak). Here, a high-spin or low-spin (two-orbital) doublon is created by the RIXS pulse and converted into a same-orbital doublon via a second-order hopping process with a neighboring doublon. In this case, energy is gained from local state transitions (Hund de-excitations) of photo-induced doublons. The gain features with $\omega_\text{out}-|E_\text{core}|\approx 8$ on the one hand originate from triplon assisted processes. In this case, a doublon is created by the RIXS pulse near $\omega_\text{in}\approx -1$, $1$, $5$ and converted into a triplon by the hopping of a neighboring triplon. On the other hand, RIXS pulses can produce triplons from low-spin doublons at $\omega_\text{in}\approx 2$, $6$. The decay of the triplon into a high-spin doublon plus filled core hole emits radiation at $\omega_\text{out}-|E_\text{core}|\approx 8$. Note that there are no apparent gain features associated with the peak at $\omega_\text{out}-|E_\text{core}|\approx 9$. Such features would require the generation of quadruplons as intermediate states, $(2,2) \xrightarrow{\omega_\text{in}} (3,1) \xrightarrow{\text{hop}} (4,1) \xrightarrow{\omega_\text{out}} (3,2)$, which is very unlikely if the density of photo-doped quadruplons is small. 

Finally, we illustrate in the bottom panels of Fig.~\ref{fig_quarter_diff} the ultrafast evolution of the RIXS signal after the photo-doping pulse in the quarter filled system. The left panel shows the difference between the signals measured with probe pulses centered at $t_\text{probe}=4.5$ and $3.5$, and the right panel the difference of the signals for $t_\text{probe}=5.5$ and $4.5$. We see that the quasi-elastic signals at $\omega_\text{in,out}\approx -2$ and $9$ increase, which indicates that the probability of holes and triplons in the system continues to increase after the pump at least until $t\approx 4.5$. This is directly confirmed in the top panel, which plots the evolution of the triplon population (the population of holes similarly increases). Also the quasi-elastic features associated with high-spin ($U-3J$) and low-spin both orbital ($U-2J$) doublons at $\omega_\text{in,out}\approx 8$ and $6$ gain some weight. This observation is again consistent with the direct measurement of the corresponding doublon populations in the upper panel. On the other hand, the quasi-elastic signal and the gain and loss features near $\omega_\text{in}\approx 2$ are suppressed, which indicates that RIXS processes involving low-spin same-orbital ($U$) doublons become less likely with increasing pump-probe delay. Indeed, these are the doublons created predominantly by the resonant pump pulse with $\omega_\text{pump}=U$, and they decay into the lower-energy doublons through scattering events (see red line in the upper panel). This example shows how the analysis of the time-resolved RIXS signal allows to extract information on the evolution of the different local state populations during and after the photo-excitation.

\section{Conclusions}

We have calculated the equilibrium and nonequilibrium RIXS spectra of the half-filled and quarter-filled two-orbital Hubbard model using a nonequilibrium-DMFT based formalism. The main idea behind this scheme is to avoid the computation and storage of four-time correlation functions by explicitly simulating the RIXS excitation pulse and measuring the self-energy of the emitted photons. This paper provides a proof-of-principle and demonstrates the feasibility of this approach in a multi-orbital context. We showed that in equilibrium our method captures $d$-$d$ excitations as well as fluorescent lines, similar to the DMFT based studies employing a configuration-interaction solver and the Kramers-Heisenberg formalism \cite{Hariki2018,Hariki2019}. After a strong photo-excitation, we observed additional peaks in the RIXS signal associated with the presence of photo-doped charge carriers, as well as new or modified gain and loss features connected to intermediate states and transitions involving these photo-carriers. We furthermore demonstrated how time-dependent changes in the RIXS spectrum can be tracked by shifting the probe pulse relative to the pump and how they reveal the evolution of different local states.

Our current implementation lacks momentum resolution and involves a certain number of approximations. For example, the core-hole lifetime is controlled by an electron bath coupled to the core level. As shown in Ref.~\cite{Eckstein2020}, this is essentially equivalent to the use of the life-time parameter $\Gamma$ in previous equilibrium \cite{Hariki2018,Hariki2019} and  nonequilibrium \cite{Chen2019,Wang2019} RIXS studies. Potentially more severe is the use of the NCA in the solution of the DMFT impurity model and the calculation of the photon self-energy. While qualitatively correct results can be expected in the Mott regime\cite{Eckstein2020}, systematic tests with beyond-NCA approaches will be needed to fully understand the implications of this approximation on the RIXS signal. It is also important to point out that while we have used here an implementation with explicit core level and classical light, there also exist alternative nonequilibrium DMFT implementations where the core level is shifted to $E_\text{core}=-\infty$, or where the outgoing photons are treated quantum mechanically \cite{Eckstein2020}. Testing these alternative implementations on multi-orbital Hubbard models will be an interesting topic for future investigations.   

\acknowledgments

We thank C. Monney, S. Johnson and H. Strand for helpful discussions. P.~W. acknowledges support from ERC Consolidator Grant 724103 and M. E. from ERC Starting Grant No.~716648. S.~J. acknowledges support from the National Science Foundation under Grant No.~DMR-1842056. The calculations were run on the beo04 cluster at the University of Fribourg.

\bibliographystyle{eplbib}

\begin{thebibliography}{99}

\bibitem{Luuk2011} Ament L., van Veenendaal M., Devereaux T., Hill J., and van den Brink J., Resonant inelastic x-ray scattering studies of elementary excitations, {\it Rev. Mod. Phys.} {\bf 83}, 705 (2011).
\bibitem{Hill1998} Hill J., Kao C.-C., Caliebe W., Matsubara M., Kotani A., Peng J., and Greene P., Resonant Inelastic X-Ray Scattering in Nd$_2$CuO$_4$, {\it Phys. Rev. Lett.} {\bf 80}, 4967 (1998).
\bibitem{Braicovich2009} Braicovich L., Ament L. J. P. , Bisogni V., Forte F., Aruta C., Balestrino G., Brookes N. B., De Luca G. M., Medaglia P. G., Miletto Granozio F., Radovic M., Salluzzo M., van den Brink J., and Ghiringhelli G., Dispersion of Magnetic Excitations in the Cuprate La$_2$CuO$_4$ and CaCuO$_2$ Compounds Measured Using Resonant X-Ray Scattering, {\it Phys. Rev. Lett.} {\bf 102}, 167401 (2009).
\bibitem{Schlappa2018}
Schlappa, J., Kumar, U., Zhou, K. J., Singh, S., Mourigal, M., Strocov, V. N., Revcolevschi, A., Patthey, L., R{\o}nnow, H. M. , Johnston, S. and T. Schmitt, T., Probing multi-spinon excitations outside of the two-spinon continuum in the antiferromagnetic spin chain cuprate Sr$_2$CuO$_3$. {\it Nat. Commun.} {\bf 9}, 5394 (2018).
\bibitem{Chen2010}
Chen, C. C., Moritz, B., Vernay, F., Hancock, J. N., Johnston, S., Jia, C. J., Chabot-Couture, G., Greven, M., Elfimov, I., Sawatzky, G. A., and Devereaux, T. P.,  
Unraveling the Nature of Charge Excitations in La$_2$CuO$_4$ with Momentum-Resolved Cu $K$-Edge Resonant Inelastic X-Ray Scattering. {\it Phys. Rev. Lett.} {\bf 105}, 177401 (2010).
\bibitem{Schlappa2012} 
Schlappa J., Wohlfeld, K., Zhou, K. J., Mourigal, M., Haverkort, M. W., Strocov, V. N., Hozoi, L., Monney, C., Nishimoto, S., Singh, S., Revcolevschi, A., Caux, J.-S., Patthey, L., R{\o}nnow, H. M., van den Brink, J. and T. Schmitt, T.,   
Spin-orbital separation in the quasi-one-dimensional Mott insulator Sr$_2$CuO$_3$, {\it Nature} {\bf 485}, 82 (2012). 
\bibitem{Ament2011} Ament L. J. P., van Veenendaal M., and van den Brink J., 
Determining the electron-phonon coupling strength from Resonant Inelastic X-ray Scattering at transition metal $L$-edges, {\it Europhys. Lett.} {\bf 95}, 27008 (2011).
\bibitem{Johnston2016}
Johnston, S., Monney, C., Bisogni, V., Zhou, K.-J., Kraus, R., Behr, G., Strocov, V. N., M{\'a}lek, J., 
Drechsler, S.-L., Geck, J., Schmitt, T. and van den Brink, J. Electron-lattice interactions strongly renormalize the charge-transfer energy in the spin-chain cuprate Li$_2$CuO$_2$, {\it Nat. Commun.} {\bf 7}, 10563 (2016).
\bibitem{Chaix2017} 
Chaix L., Ghiringhelli, G., Peng, Y. Y., Hashimoto, M., Moritz, B., Kummer, K., Brookes, N. B., He, Y., Chen, S., Ishida, S., Yoshida, Y., Eisaki, H., Salluzzo, M., Braicovich, L., Shen, Z.-X., Devereaux, T. P. and Lee, W.-S., 
Dispersive charge density wave excitations in Bi$_2$Sr$_2$CaCu$_2$O$_{8+\delta}$, {\it Nat. Phys.} {\bf 13}, 952 (2017). 
\bibitem{Dean2016} 
Dean M. {\it et al.}, 
Ultrafast energy- and momentum-resolved dynamics of magnetic correlations in the photo-doped Mott insulator Sr$_2$IrO$_4$, {\it Nature Mat.} {\bf 15}, 601 (2016). 
\bibitem{Mitrano2019} Mitrano M. {\it et al.}, Ultrafast time-resolved x-ray scattering reveals diffusive charge order dynamics in La$_{2-x}$Ba$_x$CuO$_4$, {\it Science Advances} {\bf 5}, eaax3346 (2019). 
\bibitem{Mitrano2020}
Mitrano, M. and Wang, Y., Light-Driven Quantum Materials with Ultrafast Resonant Inelastic X-Ray Scattering. 
{\it Commun. Phys.} {\bf 3}, 184 (2020). 
\bibitem{Kramers1925} Kramers H. and Heisenberg W., \"Uber die Streuung von Strahlung durch Atome, {\it Z. Phys.} {\bf 31}, 681 (1925). 
\bibitem{Chen2019} Chen Y., Wang Y., Jia C., Moritz B., Shvaika A. M., Freericks J., and Devereaux T., Theory for time-resolved resonant inelastic x-ray scattering, {\it Phys. Rev. B} {\bf 99}, 104306 (2019). 
\bibitem{Georges1996} Georges A., Kotliar G., Krauth W., and Rozenberg M. J., Dynamical mean-field theory of strongly correlated fermion systems and the limit of infinite dimensions. Dynamical mean-field theory of strongly correlated fermion systems and the limit of infinite dimensions, {\it Rev. Mod. Phys.}, 130 {\bf 68} (1996).
\bibitem{Hariki2018} Hariki A., Winder M., and Kunes J., Continuum Charge Excitations in High-Valence Transition-Metal Oxides Revealed by Resonant Inelastic X-Ray Scattering, {\it Phys. Rev. Lett.} {\bf 121}, 126403 (2018). 
\bibitem{Hariki2019} Hariki A., Winder M., Uozumi T., and Kunes J., LDA+DMFT
 approach to resonant inelastic x-ray scattering in correlated materials, {\it Phys. Rev. B} {\bf 101}, 115130 (2020).
\bibitem{Ghiringhelli2009} Ghiringhelli G., Piazzalunga A., Dallera C., Schmitt T., Strocov V., Schlappa J., Patthey L., Wang X., Berger H., and Grioni M., Observation of Two Nondispersive Magnetic Excitations in NiO by Resonant Inelastic Soft-X-Ray Scattering, {\it Phys. Rev. Lett.} {\bf 102}, 027401 (2009).
\bibitem{Wang2019} Wang Y., Chen Y., Jia C., Moritz B., and Deveraux T., Time-Resolved Resonant Inelastic X-Ray Scattering in a Pumped Mott Insulator, {\it Phys. Rev. B} {\bf 101}, 165126 (2020). 
\bibitem{Eckstein2020} M. Eckstein and P. Werner, Simulation of time-dependent resonant inelastic X-ray scattering with non-equilibrium dynamical mean-field theory, {\it arxiv} (2020). 
\bibitem{Aoki2014} Aoki H., Tsuji N., Eckstein M., Kollar M., Oka T., and Werner P., Nonequilibrium dynamical mean-field theory and its applications, {\it Rev. Mod. Phys.} {\bf 86}, 779 (2014).
\bibitem{Keiter1971} Keiter H. and Kimball J. C., A revised diagrammatic technique for the degenerate Anderson model, {\it Int. J. Magn.} {\bf 1}, 233 (1971).
\bibitem{Eckstein2010} Eckstein M. and Werner P., Nonequilibrium dynamical mean-field calculations based on the noncrossing approximation and its generalizations, {\it Phys. Rev. B} {\bf 82}, 115115 (2010).
\bibitem{footnote}{A realistic core level would be at much lower energy. Here, we choose a shallow core level to enable calculations with a time step $\Delta t=0.01$ that nevertheless resolve the short timescale $\hbar/|E_\text{core}|$. For larger $E_\text{core}$, one should write the core-valence Hamiltonian in the rotating wave approximation~\cite{Eckstein2020}. However, already for  $E_\text{core}=20$, counter-rotating terms give only a small contribution to the result. 
}
\bibitem{Lysne_2020} Lysne M., Murakami Y. and Werner P., Signatures of bosonic excitations in high-harmonic spectra of Mott insulators, {\it Phys. Rev. B} {\bf 101}, 195139 (2020).
\bibitem{Strand_2017} Strand H., Golez D., Eckstein M., and Werner P., Hund's coupling driven photocarrier relaxation in the two-band Mott insulator, {\it Phys. Rev. B} {\bf 96}, 165104 (2017).
\bibitem{Eckstein_2011} Eckstein M. and Werner P., Thermalization of a pump-excited Mott insulator, {\it Phys. Rev. B} {\bf 84}, 035122 (2011).
\bibitem{Petocchi_2019} Petocchi F., Beck S., Ederer C., and Werner P., Hund excitations and the efficiency of Mott solar cells, {\it Phys. Rev. B} {\bf 100}, 075147 (2019).


\end{thebibliography}

\end{document}